# TAKING A MOMENT TO MEASURE NETWORKS – A HIERARCHICAL APPROACH

**Word Count: 8841**

**Kehinde R. Salau[1], Jacopo A. Baggio[2], Marco A. Janssen[3], Joshua K. Abbott[3], Eli P. Fenichel[4]**

[1]Department of Mathematics, University of Arizona, 617 N Santa Rita Ave, Tucson, AZ 85721, USA. **Email**: krsalau@email.arizona.edu

[2] Department of Environment and Society, Utah State University, 5215 Old Main Hill, Logan, UT 84322-5215, USA

[3]Global Institute of Sustainability, School of Sustainability, Arizona State University, 800 S Cady Mall, Tempe, AZ 85281, USA

[4]Yale School of Forestry and Environmental Studies, Yale University, 195 Prospect St, New Haven, CT 06511, USA

**ABSTRACT**

Network-theoretic tools contribute to understanding real-world system dynamics, e.g., in wildlife conservation, epidemics, and power outages. Network visualization helps illustrate structural heterogeneity; however, details about heterogeneity are lost when summarizing networks with a single mean-style measure. Researchers have indicated that a hierarchical system composed of multiple metrics may be a more useful determinant of structure, but a formal method for grouping metrics is still lacking. We develop a hierarchy using the statistical concept of moments and systematically test the hypothesis that this system of metrics is sufficient to explain the variation in processes that take place on networks, using an ecological systems example. Results indicate that the moments approach outperforms single summary metrics and accounts for a majority of the variation in process outcomes. The hierarchical measurement scheme is helpful for indicating when additional structural information is needed to describe system process outcomes.

**INTRODUCTION**

Network theory is ubiquitous across the applied sciences (Boccaletti et al. 2006; Barthélemy 2011; Blonder et al 2012). Networks are appealing because they provide clear visualizations of interlinked systems, and networks preserve heterogeneities and local information. The motivating hypothesis implicit in network analysis is that by understanding the underlying structure of linkages, researchers gain predictive power about processes taking place on networks, e.g., the dispersal and persistence of organisms (Urban et al. 2009), infectious disease dynamics (May 2006), neuron communication (Laughlin & Sejnowski 2003), and the diffusion of ideas (Watts 2002). Networks are often described using summary statistics such as mean degree, mean shortest path, and mean clustering coefficient (Estrada & Bodin 2008). Summary statistics give an overview of the network linkages, but the relationship between summary statistics and processes is unclear *ex ante*. Furthermore, details about heterogeneity vanish when summarizing networks with a single mean-style metric. A hierarchical system composed of multiple metrics could aid research in the analysis of network structures, but a formal method for grouping network metrics is lacking (Estrada & Bodin 2008). We fill this gap in the literature and develop a hierarchy of network metrics and systematically test the hypothesis that simple metrics suffice to explain the variation in processes playing out on networks. The nested hierarchy of metrics is motivated by the statistical concept of moments, where a set of numerical features are systematically calculated and used to describe the structure of a distribution—or, in the case of a network, a set of connections among nodes—in increasing cumulative detail.

Network science is awash with approaches for measuring networks. Barrat et al. (2004) use the mean clustering coefficient, a measure of local cohesion defined by node degree and edge weights, to study the effects of topology and node interaction strength in a scientific collaboration network and the worldwide air-transportation network. Liu et al. (2013) use global efficiency, the inverse of the harmonic mean of the total number of pairwise shortest paths, to parse the effects of Alzheimer's disease on human brain networks. Rayfield, Fortin, & Fall (2011) highlight the popularity of summary indices in ecology, asserting that the number of publications using network theory to quantify habitat networks has grown tenfold over the past three decades. Many established metrics for measuring network connectivity are strongly correlated (Baggio et al. 2011), but a clear hierarchy is lacking.

The spectral radius of a matrix of edge weights is a fundamental measure in the analysis of social, biological, and infrastructure networks (van Mieghem 2011). The spectral radius faces the same limitations as any other single metric because it summarizes global network structure. However, derivation of the spectral radius also yields the eigenvector centrality, which normalizes the information on all the linkages in a network. Though it preserves a great deal of local information, a drawback of eigenvector centrality is that it does not provide a simple summary statistic. We use spectral radius and eigenvector centrality, collectively known as eigenmetrics, to demonstrate a hierarchical approach to measuring networks. Specifically, we apply the concept of moments by treating the eigenvector centrality as a distribution of node connectivity scores. Different moments (e.g. mean, variance, skewness) of the resulting distribution highlight different topological properties of networks; the interplay among these network "moments" is useful for describing, and potentially predicting, processes occurring on networks.

In this paper we present a general hierarchical approach to evaluate the impact of network structure on outcomes. First, we outline the theoretical underpinnings of the approach. Then, we demonstrate the method by applying it to the study of prairie dog metapopulation dynamics. Prairie dog population growth is largely driven by variable, individual dispersal to spatially distinct prairie towns, so a simple mean field model may fail to capture important local information (Durrett & Levin 1994). We use an agent-based modeling approach to capture such dynamics over a network of prairie dog towns. Agent-based models (ABMs) are widely used in relevant studies on individual behavior, spatial population dynamics and conservation (Grimm & Railsback 2005; West et al. 2011; Sibly et al. 2013; Schoon et al. 2014). Our results demonstrate the potential for the hierarchical approach to be a standard method for grouping networks and parsing outcomes, but more work needs to be done to assess the approach on a broader set of applications.

**MATERIALS AND METHODS**

**Measuring a network by moments**

Consider a network $G$ with $N$ nodes, where each pair of nodes is connected by a weighted edge that represents the relative ease of movement or information spread through the network, with lower weights leading to less resistance on the network and easier movement. $G$ can be expressed as an $N \times N$ adjacency matrix, denoted $A_G$, where the edge weights between the $N$ nodes of $G$ make up the elements of $A_G$ (figure 1). $A_G$ is always a zero-diagonal matrix, as information faces no resistance to stay at a node. In an ecological context landscape networks and metapopulation models account for changes within a node through local birth, mortality, and predation events (Hanski & Gilpin 1997). Edge direction can play a substantial role on network dynamics especially when dealing with issues of asymmetry (e.g. uphill/downhill transportation, unreciprocated contact, etc.). We develop the hierarchical framework in the context of bidirectional networks, which are common in network science (Urban & Keitt 2001; Boit et al. 2012), but it can be generalized to directional networks by modeling inflows and outflows as separate edges.

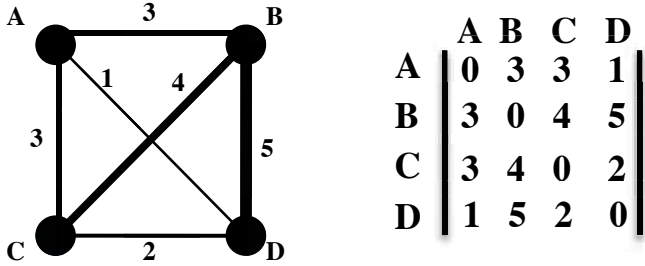

*Figure 1*. (Left) A weighted 4-node network denoted G. The weights, denoted numerically and by edge width, act as facilitators/inhibitors of movement along some dispersal corridor between nodes. (Right) The corresponding adjacency matrix $A_G$.

The spectral radius, $\lambda_G$, is the dominant eigenvalue of $A_G$ and measures the overall traversability or mean distance across a network (Jacobi & Jonsson 2011). A network with low spectral radius is less resistant and highly connected. An increase in the spectral radius indicates a decrease in connectivity. Spectral radius is a mean measure, so information is lost when it is used to summarize network characteristics. This may be acceptable for some analysis, but unacceptable for others. For a network with a given number of nodes and weighted edges, there is an infinite set of network configurations for any spectral radius, and these different configurations can lead to different process outcomes. This problem is not unique to spectral radius. For example, many different disease outcomes are possible on networks with the same mean degree (May 2006).

The adjacency matrix can also be used to recover the eigenvector centrality (EC) of $G$, which describes the importance of an individual node within a network. The EC is the $N \times 1$ eigenvector ($\vec{v}_G$) associated with the spectral radius whose elements are rescaled so the Euclidean norm of $\vec{v}_G$ is 1. The $i^{th}$ component of the EC ranks the importance of the $i^{th}$ node as donor and recipient of information within the network and describes its contribution to network connectivity (Urban et al. 2009). A node with a low EC score is highly connected relative to other nodes in the network. So the EC provides a value for each node, but this does not help summarize the network. To summarize the EC, we treat the elements of an $N$-dimensional EC as $N$ data points and use the statistical moments of the corresponding empirical distribution. However, we discard the mean, the first moment, because there is a one-to-one relationship between EC mean and EC variance, the second moment (*SI Text S1*).

Variance measures the spread in a dataset. In the network context, EC variance ($var(\vec{v}_G)$) measures the spread in node contribution across the network and provides a measure of heterogeneity among nodes. Zero EC variance implies that all nodes contribute equally to global connectivity (figure 2A). Networks with nonzero EC variance contain at least two nodes that contribute unequally (figure 2B).

Skewness, the third moment, indicates whether deviations from the mean of a dataset are systematically positive or negative and measures the level of asymmetry in data. We normalize the deviations with variance when calculating skewness. Datasets with negative skew contain a larger proportion of points exceeding the mean. EC skewness ($skew(\vec{v}_G)$) captures the net ratio of relatively strong to weak contributors. Node *i* of network *G* is a relatively strong (weak) contributor if its corresponding EC score ($v_i \in \vec{v}_G$) is less (greater) than the EC mean. Networks with negative EC skewness possess a larger proportion of weak contributors (figure 3A), zero EC skewness reflects a one-to-one ratio of weak to strong contributors and structures with positive EC skewness have a higher proportion of strong contributors (figure 3B). Our approach of using EC moments could be extended to higher order moments, but it is hard to produce clear interpretable meaning for statistical moments past the third (Casella & Berger 2002).

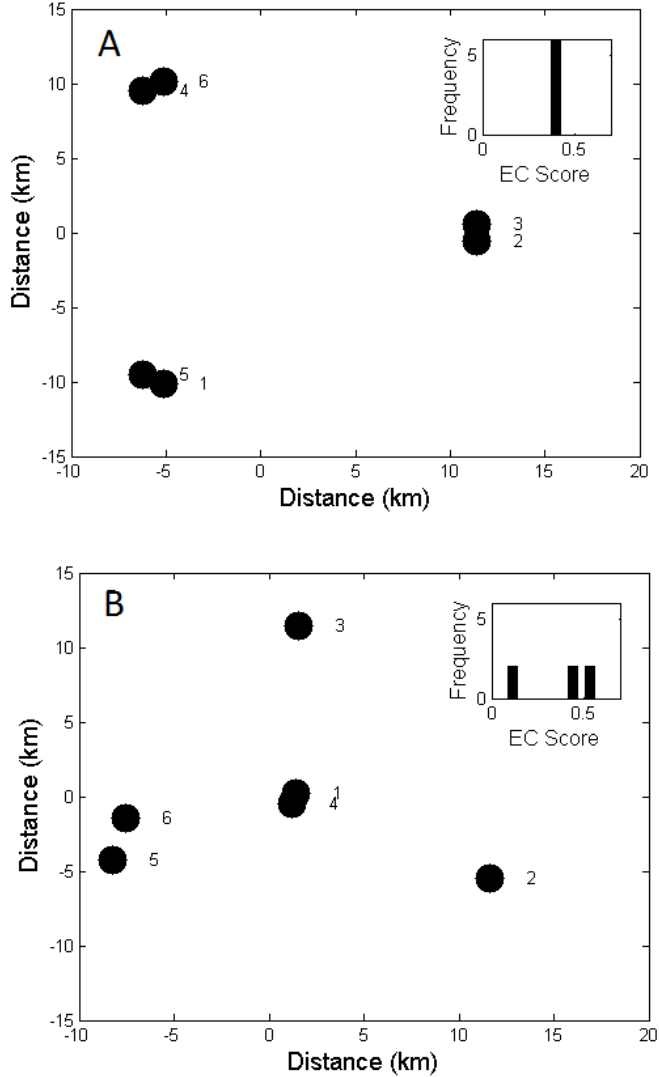

*Figure 2*. Two 6-node networks with the same spectral radius ($\lambda_G = 80km$) but different EC variance. (A) Network with zero EC variance. (B) Network with nonzero EC variance ($var(\vec{v}_G) = 0.026$). Inset A and B contain the frequency distribution of EC scores for nodes in networks (A) and (B) respectively. Nodes 1 and 4 in Network (B) are more connected than other nodes hence node contribution is not homogeneous.

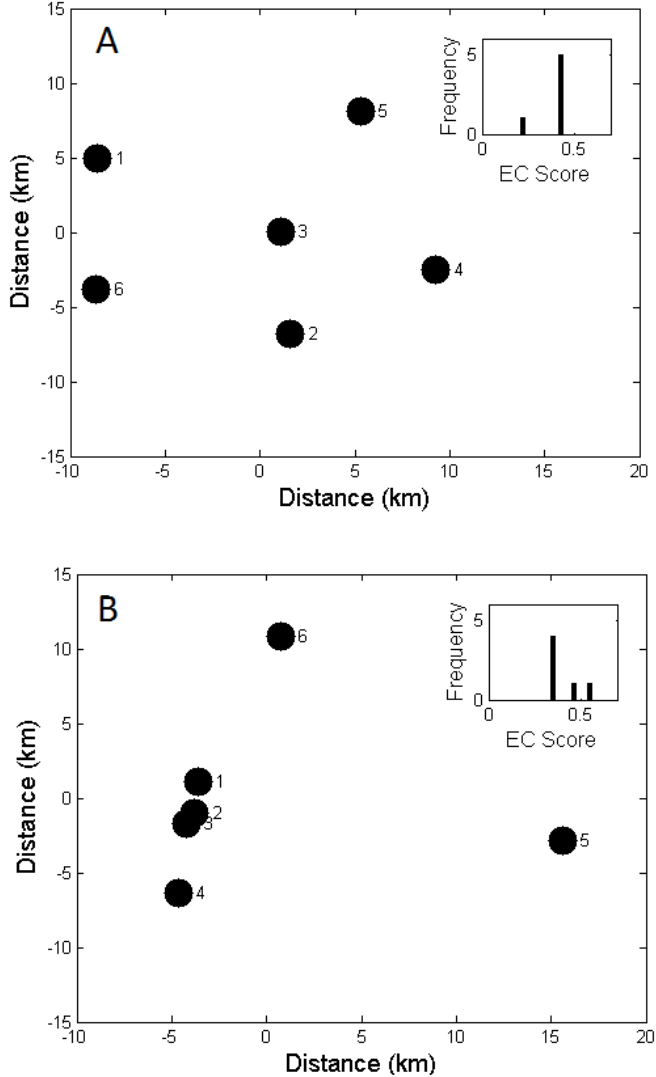

*Figure 3*. Two 6-node networks with the same spectral radius ($\lambda_G = 65km$) and EC variance ($var(\vec{v}_G) = 0.0086$) but different EC skewness. (A) Network with negative EC skewness ($skew(\vec{v}_G) = -1.79$). (B) Network with positive EC skewness ($skew(\vec{v}_G) = 1.086$). Node 3 in Network (A) is the only strong contributor on the landscape. In contrast, nodes 1-4 in Network (B) are strong contributors.

**Connecting eigenmetrics with other popular network metrics**

Spectral radius is positively correlated with the mean strength and the mean clustering coefficient of a network. We use a rescaling argument to derive the mathematical relationship between spectral radius and mean strength and then relate mean strength to mean clustering coefficient by a constant factor; the latter exercise implies a connection between spectral radius and mean clustering coefficient via transitivity (see *SI Text S2* for derivations). EC variance is closely related to common metrics that are not highly related to spectral radius, e.g. mean shortest path length, global efficiency, and local efficiency. This supports the idea that statistical moments avoid redundancy and are useful for organizing information from a large set of available summary indices. We show that mean shortest path length characterizes a strong lower bound for EC variance and then use a similar transitivity argument to link global and local efficiency with EC variance (*SI Text S2*). Figure 4 summarizes the mathematical connections among the metrics discussed above.

**Model design**

A common use of network analysis is the measurement of habitat connectivity for species conservation (Urban & Keitt 2001; Dixon et al. 2006). In the absence of extensive data on the ecology of species and interactions with the landscape, model simulation is a useful tool for analyzing the ecological implications of landscape structure (Urban et al. 2009; Moilanen 2011; Rebaudo et al. 2013). We use a model of animal movement on a physical landscape to limit the variability of network structure and illustrate our hierarchical approach. In the network formulation, nodes represent habitat patches and edges represent corridors that facilitate individual dispersal. A desirable feature of a general hierarchical approach is that it is robust to multiple processes. Therefore, we consider outcome variation in two ecological processes: spread and survival. Spread potential is measured as the time needed for an initial population on one randomly chosen node to occupy the last uninhabited node (i.e. time to full network occupation). Survival is measured as time to global extinction (i.e. no individuals on any patch).

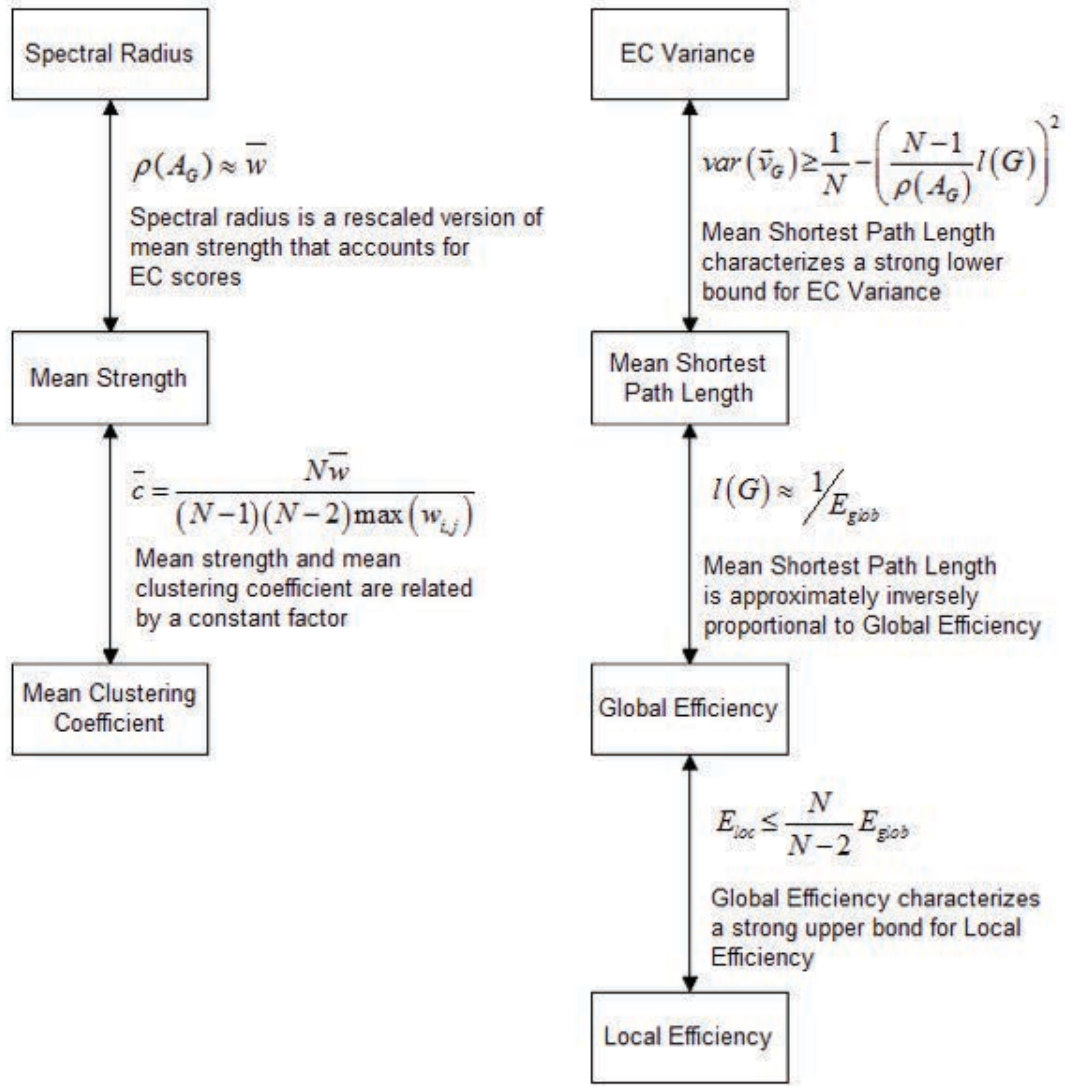

*Figure 4*. Relating eigenmetrics to other popular network metrics. Many of the above relationships result from the fact that adjacency matrices for the class of network we study are fully connected, zero-diagonal, nonnegative and symmetric. $\bar{w}$ = mean strength, $E_{glob}$ = global efficiency, $l(G)$ = mean shortest path length, $E_{loc}$ = mean local efficiency, $\bar{c}$ = mean clustering coefficient.

The data are drawn from agent-based representations of single-species habitat networks (described in the SI). The actions of the agents are probabilistic and provide a scenario where the predictive power of the chosen metrics is assessable amid stochastic population dynamics. The ABMs are calibrated using data on prairie dogs (*Cynomys* spp.). Prairie dog metapopulations can be represented as a network of multiple complexes, consisting of multiple prairie dog families, with low-lying drainages, roadways and other landscape features serving as corridors (Roach et al. 2001). Distance is a dominating factor in successful prairie dog dispersal along drainages and roadways, supporting the assumption of symmetry (Garrett & Franklin 1988; Bevers et al. 1997; Holmes 2008).

We develop an algorithm (outlined in *SI Text S3*) to build 6×6 adjacency matrices with predetermined spectral radius, EC variance, and EC skewness (in MATLAB R2012a). The matrices represent 6 connected nodes, which is comparable to networks of prairie dog complexes (Antolin, Savage & Eisen 2006). An ABM of prairie dogs (implemented in NetLogo 5.0.1) is simulated on these constructed networks. On average prairie dogs disperse 4km (Bevers et al. 1997). 20km is used as the upper bound for any edge weight reflecting a low probability of successful dispersal (Holmes 2008). The minimum distance between nodes is 1km, corresponding to the minimum distance between prairie dog complexes (Holmes 2008). If all nodes were 1km apart, then the network's spectral radius is 5km. Conversely, if all nodes where 20km apart, then the network has a spectral radius of 100km. Networks with equal weights on all edges have zero EC variance, resulting in an undefined EC skewness. We choose 5 spectral radii, spanning the spectrum of potential spectral radii in our system, to generate adjacency matrices. For each spectral radius we specify five EC variance measures, and then repeat the process with five levels of EC skewness. After accounting for networks that have an EC variance of 0, there are 107 networks configurations. We simulate 400 realizations per network configuration for each ecological process.

**Local dynamics**

The ABM birth, death, and dispersal events are stochastic. Prairie dogs exhibit density-dependent growth (Hoogland et al. 1988). In each time-step, a prairie dog on node $i$ produces $f_x$ offspring with probability, $1 - exp[-r(1 - D_{x,i})]$; $r$ is the intrinsic growth rate of prairie dogs. $D_{x,i}$ denotes prairie dog density on node $i$ and is computed as $D_{x,i} = x_i/K_i$, where $x_i$ is the absolute number of prairie dogs and $K_i$ represents prairie dog carrying capacity for node $i$. Prairie dog mortality on node $i$ occurs with probability $q_x$. Table 1 provides a summary of agent attributes and parameters.

**Dispersal dynamics**

We follow prior analytical and computational models of dispersal dynamics (Amarasekare 2004; Tang & Bennett 2010), and divide dispersal into the decision to disperse and the likelihood of successful dispersal. Intraspecific competition influences prairie dog dispersal (Hof et al. 2002). In the ABM, prairie dogs disperse from node $i$ with density-dependent probability,

| Symbol | Description | Value |
|---|---|---|
| $N_x$ | Initial number of prairie dogs on a patch | 150 |
| $r$ | Prairie dog growth rate | 0.74[a] |
| $f_x$ | Prairie dog litter size | 3[b] |
| $q_x$ | Prairie dog mortality probability | 0.4[a] |
| $D_{U,x}$ | Prairie dog density threshold affecting own dispersal | 0.9[c] |
| $M_x$ | Average dispersal distance of prairie dogs | 2km[d] |
| $K_i$ | Prairie dog carrying capacity on patch $i$ ($i = 1, 2$) | 150 |

*Table 1*. Summary of variables and parameters used in the ABM. Approximated from: [a](Klebanoff et al. 1991), [b](Hoogland et al. 1988), [c](Salau et al. 2012), [d](Garrett & Franklin 1988). The prairie dog parameters are compiled from several different regions and are intended to bound the parameter space, not outline a specific case study.

$$\begin{cases} D_{x,i} / D_{U,x} & \text{if} \quad D_{x,i} < D_{U,x} \\ 1 & \text{if} \quad D_{x,i} \geq D_{U,x} \end{cases} \quad [1]$$

$D_{U,x}$ is a fixed density threshold indicator of overcrowding below which, the decision to disperse is random but increasingly likely with higher prairie dog density. Above $D_{U,x}$, dispersal is certain.

Assuming an individual animal disperses out of a node, the probability of successful arrival at another node is a function of distance (Hof et al. 2002) and inversely related to the edge weight between two nodes. A dispersing animal completes a move from node $i$ to node $j$ if,

$$Exp(M_x) < W_{ij} \quad [2]$$

The term $Exp(M_x)$ represents a random variate drawn from the exponential distribution with mean $M_x$, which denotes the mean dispersal ability of prairie dogs. The edge weight $W_{ij}$ is the corridor distance between nodes $i$ and $j$. If the animal cannot reach node $j$ given the dispersal ability draw, then the dispersing animal dies. A description of the sequence of agent events for each ecological process is available in *SI Text S4* and archived online (www.openabm.org/model/4621/version/1/view).

**Results**

**Single metrics**

Single metrics collapse the complex system into a single dimension allowing for coarse comparisons. Therefore, we investigate the relationship between single metrics and ecological scenarios in order to describe the basic trend between network structure and function. High spectral radius represents low traversability, which limits successful dispersal through the landscape (figure 5A). Spread is faster in networks with high EC variance because structures with greater node heterogeneity contain a strongly connected node that, once inhabited, facilitates spread to all nodes (figure 5B). EC skewness does not have a clear relationship to spread (figure 5C). We also capture similar trends using other common metrics; the resulting metric correlations lend support to the mathematical derivations in the previous section (see Table 2).

A prominent working hypothesis in conservation is that connectivity is important for conservation; this is the rationale for maintaining connectivity between habitat patches (Hanski & Gilpin 1997). Our simulations support this claim. Networks with low spectral radius coincide with longer persistence times (figure 5D). Greater connectivity, indicated by lower spectral radius, allows greater mobility for foraging, securing refuge, and re-colonization. Networks with greater EC variance coincide with longer persistence times; in this case, the strongly connected node is the source of re-colonization and provides a rescue effect (figure 5E). All else equal, structures consisting of a larger proportion of strong contributors (i.e. high EC skewness) support longer survival periods.

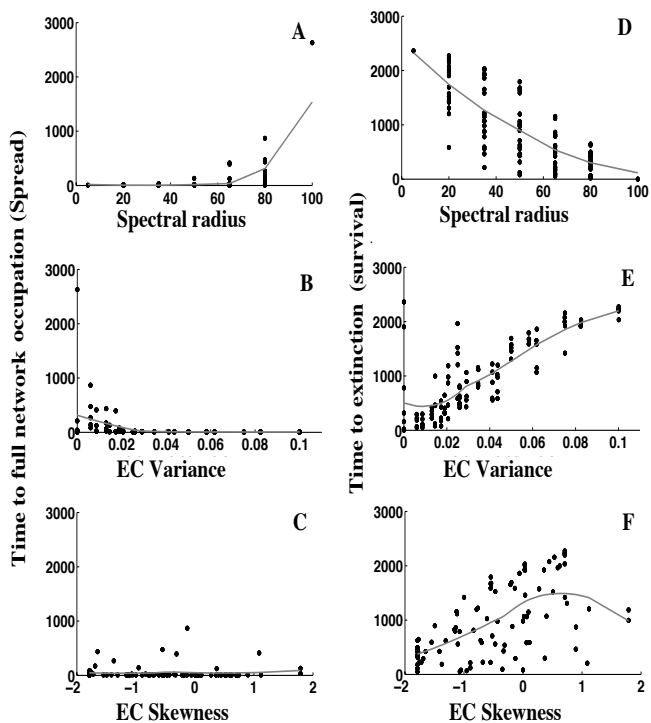

*Figure 5. Relating single indicators to ecological function. The columns of plots each pertain to the spread and survival scenario respectively. The y-axis on each plot represents a median time measure. The solid line in each plot represents the trend in the data given by LOWESS estimation (see SI Text S5 for details on the calculation). (A-B) Extremely slow spread times for networks with high spectral radius ($\lambda_G = 100km$) or low EC Variance ($var(\vec{v}_G) = 0$) suggest that spread regression models involving these metrics may better explain spread rate when plotted against log spread time.*

Node heterogeneity, measured by EC variance, strongly influences network traversability (spectral radius) in all ecological scenarios, suggesting tradeoffs among distinct structural properties. Understanding and measuring these multiple properties is likely important for conservation planning and requires multiple metrics. Ames et al. (2011) make a similar argument for disease dynamics on networks.

|  | $\lambda_G$ | $avg(\vec{v}_G)$ | $var(\vec{v}_G)$ | $skew(\vec{v}_G)$ | $\bar{w}$ | $E_{glob}$ | $l(G)$ | $E_{loc}$ | $\bar{c}$ |
|---|---|---|---|---|---|---|---|---|---|
| $\lambda_G$ | 1.00 | | | | | | | | |
| $avg(\vec{v}_G)$ | 0.68 | 1.00 | | | | | | | |
| $var(\vec{v}_G)$ | -0.67 | -0.99 | 1.00 | | | | | | |
| $skew(\vec{v}_G)$ | -0.75 | -0.47 | 0.46 | 1.00 | | | | | |
| $\bar{w}$ | 0.98 | 0.80 | -0.80 | -0.74 | 1.00 | | | | |
| $E_{glob}$ | -0.71 | -0.90 | 0.90 | 0.52 | -0.82 | 1.00 | | | |
| $l(G)$ | 0.72 | 0.92 | -0.92 | -0.40 | 0.82 | -0.94 | 1.00 | | |
| $E_{loc}$ | -0.74 | -0.90 | 0.90 | 0.57 | -0.84 | 0.99 | -0.93 | 1.00 | |
| $\bar{c}$ | 0.97 | 0.82 | -0.82 | -0.76 | 0.99 | -0.85 | 0.83 | -0.87 | 1.00 |

*Table 2*. Correlation between eigenmetrics and popular network metrics. *Spearman coefficients are all significant at 5% level. $avg(\vec{v}_G)$ = mean of the eigenvector centrality, $\bar{w}$ = mean strength, $E_{glob}$ = global efficiency, $l(G)$ = mean shortest path length, $E_{loc}$ = mean local efficiency, $\bar{c}$ = mean clustering coefficient.

**Single vs. multiple metrics: A statistical test of significance**

Despite qualitative relationship between single metrics and ecological processes outcomes, a large degree of variation remains unexplained (figure 5). One reason is that single metrics do not allow multiple structural attributes to be considered simultaneously. We measure how much variation is explained when single metrics are combined. We use regression models to investigate how different combinations of spectral radius, EC variance, and EC skewness perform as predictors of two outcomes: median time to full network occupation and median time to single-species extinction.

The greatest amount of variation in both ecological scenarios is best explained using all three metrics (Table 3). Even when penalizing models with extra predictors, models using all three metrics are still of greater quality than models using less (Table 3). Some metrics do not explain much variation as single predictors, but markedly influence fit when conditional on controlling for other metrics. For example, with an $R^2$ of zero, EC skewness is unreliable as the sole predictor of spread but increases predictive power

| Ecological process | Network metric(s) | | | R² | ΔAIC | Rank |
|---|---|---|---|---|---|---|
| **Median spread time** | $\lambda_G$ | | | 0.135 | 143.982 | 5 |
| | | $var(\vec{v}_G)$ | | 0.118 | 155.350 | 6 |
| | | | $skew(\vec{v}_G)$ | 0.000 | 230.866 | 7 |
| | $\lambda_G$ | $var(\vec{v}_G)$ | | 0.151 | 134.438 | 4 |
| | $\lambda_G$ | | $skew(\vec{v}_G)$ | 0.309 | 11.200 | 2 |
| | | $var(\vec{v}_G)$ | $skew(\vec{v}_G)$ | 0.162 | 126.968 | 3 |
| | $\boldsymbol{\lambda_G}$ | $\boldsymbol{var(\vec{v}_G)}$ | $\boldsymbol{skew(\vec{v}_G)}$ | **0.324** | **0[f]** | 1 |
| **Median survival time** | $\lambda_G$ | | | 0.573 | 590.130 | 6 |
| | | $var(\vec{v}_G)$ | | 0.797 | 144.130 | 4 |
| | | | $skew(\vec{v}_G)$ | 0.347 | 845.828 | 7 |
| | $\lambda_G$ | $var(\vec{v}_G)$ | | 0.839 | 8.222 | 2 |
| | $\lambda_G$ | | $skew(\vec{v}_G)$ | 0.575 | 589.270 | 5 |
| | | $var(\vec{v}_G)$ | $skew(\vec{v}_G)$ | 0.826 | 55.142 | 3 |
| | $\boldsymbol{\lambda_G}$ | $\boldsymbol{var(\vec{v}_G)}$ | $\boldsymbol{skew(\vec{v}_G)}$ | **0.841** | **0[g]** | 1 |

**Table 3**. *Table of sample regression models using spectral radius ($\lambda_G$), EC variance ($var(\vec{v}_G)$), and EC skewness ($skew(\vec{v}_G)$) as predictor variables and median time values from the spread and survival scenarios as response variables. $R^2$ indicates the proportion of variability in outcomes explainable by a given model. AIC provides a measure of model fit that penalizes extra predictors; preferred models have lower AIC. ΔAIC is a rescaling of original AIC values by the lowest AIC value in the group of models. Original AIC value:* [f]*7231.528,* [g]*8396.336*

when paired with the other metrics. When comparing the spread regression model of spectral radius and EC variance with the model using all three metrics, we find that the latter explains twice as much variation. In general, network metrics are relatively poor predictors of spread; an unexpected result because landscape structure is expected to directly determine dispersal but indirectly influence persistence. The $R^2$ for the spread model improves to 64 percent when squared and interaction terms are included in the regression (see Table 4).

| Ecological process | Network metrics | | | | | $R^2$ | $\Delta AIC^a$ |
|---|---|---|---|---|---|---|---|
| **Spread** | $\lambda_G$   $var(\vec{v}_G)$   $skew(\vec{v}_G)$   $\lambda_G^2$   $\lambda_G \times var(\vec{v}_G)$ $\lambda_G \times skew(\vec{v}_G)$ | | | | | 0.643 | -376.75 |
| **Survival** | $\lambda_G$   $skew(\vec{v}_G)$   $\lambda_G^2$   $var(\vec{v}_G)^2$   $skew(\vec{v}_G)^2$ $\lambda_G \times var(\vec{v}_G)$   $var(\vec{v}_G) \times skew(\vec{v}_G)$ | | | | | 0.886 | -188.97 |

***Table 4***. *Best models from regression.*
*[a]Difference in AIC between the model presented here and the best model reported in Table 3.*

**Sensitivity analysis**

Parameter choice can bias results from computational models and, in this study, hamper general claims of statistical significance. ABMs are a boon in this regard because they allow for repeated scenario testing and targeted assessment of parameter effects in a controlled environment. We perform sensitivity analysis on the population parameters of the prairie dog ABM and assess whether multi-metric regression models always outperform single metric models. We give each default parameter value a ten percent increase/decrease, collect new simulation data, and document the change in $R^2$ and AIC values for the regression models. A series of tables (one for each parameter perturbation) containing the adjusted statistical measures can be found in the supplementary material; we give further instructions on reading the tables in *SI Text S6*. For two parameter perturbations (increased prairie dog litter size and decreased mortality) in the survival scenario, extinction events became so rare in our simulation context that it was not feasible to compare network structures. This is because reproduction and survival on any one node was sufficiently high.

Though the ranking of single and two-metric regression models change depending on parameter settings, we find that the model with all three metrics always provides the best indicator of spread and survival. Within this subset of models, $R^2$ values for the spread scenario are less sensitive to perturbation than their survival counterparts (Table 5). So despite an overall lower $R^2$, the three metrics are more robust predictors of spread in a network. Robustness is key in the complete assessment of network metrics on different types of outcomes.

|  | Perturbed value | $\triangle R^2$ | |
|---|---|---|---|
| Parameter | -10% (+10%) | Spread | Survival |
| Initial number of prairie dogs on a patch, $N_x$ | * | -0.012 (-0.011) | 0.001 (-0.015) |
| Prairie dog growth rate, $r$ | 0.67 (0.82) | -0.002 (0.011) | -0.191 (-0.426) |
| Prairie dog litter size, $f_x$** | 2 (4)*** | 0.006 (0.043) | -0.339 (NA) |
| Prairie dog mortality probability, $q_x$ | 0.504*** (0.616) | -0.001 (-0.009) | NA (-0.313) |
| Prairie dog density threshold, $D_{U,x}$ | 0.81 (0.99) | -0.004 (0.002) | -0.002 (-0.008) |
| Average dispersal distance of prairie dogs, $M_x$ | 1.8km (2.2km) | -0.035 (0.031) | -0.021 (0.004) |
| Carrying capacity, $K_i$ | 135 (165) | 0.002 (0.004) | -0.036 (-0.003) |

***Table 5***. *This table provides the quantitative change in $R^2$ for the 3-metric linear regression model when perturbing model parameters. We systematically increase/decrease each default parameter by ten percent then recalculate the relationship between network metrics and outcomes. Positive $\triangle R^2$ implies that the $R^2$ associated with the perturbed model is greater than the largest $R^2$ value reported in Table 3.*
*\*These values are scenario-dependent. For spread, the perturbed values are 4 (6). For survival, the perturbed values are 135 (165).*
*\*\*These values must be nonnegative integers.*
*\*\*\*We are unable to observe any meaningful relationship between metrics and median survival time because simulations with death rate 0.504 (and lower) or litter size 4 (and higher) seldom lead to extinction.*

In the survival scenario, regression performance is highly sensitive to prairie dog growth rate and litter size parameters; in one experiment, the original $R^2$ value reduced by a factor of 2 (Table 5). But even at the lowest $R^2$ level, the 3-metric model remains a better predictor of survival than spread, which again is surprising given the presumed connection between network structure and dispersal. Perhaps this result is less astounding when one also considers the important linkage between dispersal and survival in the case of prairie dogs. Ultimately, tradeoffs in the accuracy and robustness of metrics are realistic, unavoidable, and amplify the hardships managers face when seeking to understand and influence dynamics on networks. The hierarchical approach, coupled with a controllable model, helps quantify these tradeoffs and inform the discussion on how to best summarize networks.

**Hierarchy of metrics**

Regression and sensitivity analyses support simultaneous use of multiple metrics when evaluating network processes. We sort the metrics in the same order they are derived—this is the hierarchy. The sorting helps develop a narrative on how the combined effects of the metrics dictate process outcomes. We highlight regions in the spread and survival scenario where single metrics tell an incomplete story and indicate when a single metric is sufficient (see Table 6 for a summary).

|  |  |  | EC Variance | | |
|---|---|---|---|---|---|
|  |  |  | **Low** | **Intermediate** | **High** |
|  | **Low (5-20)** | **Spread** | Fast | Fast | Fast |
|  |  | **Survival** | *Varies*[a] | *Varies*[b] | Long |
| **Spectral Radius** | **Intermediate (35-65)** | **Spread** | *Varies*[c] | Fast | Fast |
|  |  | **Survival** | *Varies*[d] | *Varies*[e] | Long |
|  | **High (80-100)** | **Spread** | *Varies*[c] | Fast | Fast |
|  |  | **Survival** | Brief | Brief | Brief |

*Table 6*. Summary of spread and survival outcomes.
[a]Survival time is highly variable, but normally long at margins of EC skewness; [b]Survival time is generally long but decreases with higher EC skewness; [c]Slower spread time for structures with higher EC skewness; [d]Survival is brief but increases with higher EC skewness; [e]Survival is brief but increases at margins of EC skewness.

The key determinant of fast spread is node accessibility, and greater node accessibility is indicated by a combination of low spectral radius and high EC variance (Table 6). Spectral radius is a sufficient indicator of fast spread in networks with moderate to high traversability (in our example, $\lambda_G \leq 50$km) due to the overall closeness of nodes. In networks with low traversability ($\lambda_G > 50$km), node heterogeneity becomes the deciding factor and greater EC variance indicates fast spread (figure 6B).

We observe variation in spread outcomes for networks with low traversability ($\lambda_G \geq 65$km) and low heterogeneity ($var(\vec{v}_G) \leq 0.01$); see region *I* (figure 6B). For such networks, spectral radius and EC variance are unable to fully capture all scenarios and EC skewness provides additional information for parsing through different spread outcomes. For networks with high EC skewness in region *I*, node accessibility is lower even though the proportion of strong contributors is high. This is because of an extremely weak contributor, which results from increasing EC skewness in region *I* (recall figure 3B). As a result, dispersing agents are unlikely to reach or escape the isolated node. Region *I* is a prime example of how complex the determination of spread outcomes becomes when using the networks concept. The ability to illustrate and explain different outcomes in region *I* is a key advantage of the hierarchical moments approach.

Node accessibility is also important for persistence. At the highest levels of spectral radius ($\lambda_G \geq 80$km), all nodes are isolated and survival is brief (figure 7B); spectral radius is a good predictor of persistence outcomes in such structures. For networks with intermediate spectral radius persistence becomes a function of node heterogeneity; all nodes are accessible and survival time increases with increased EC variance (figure 7A).

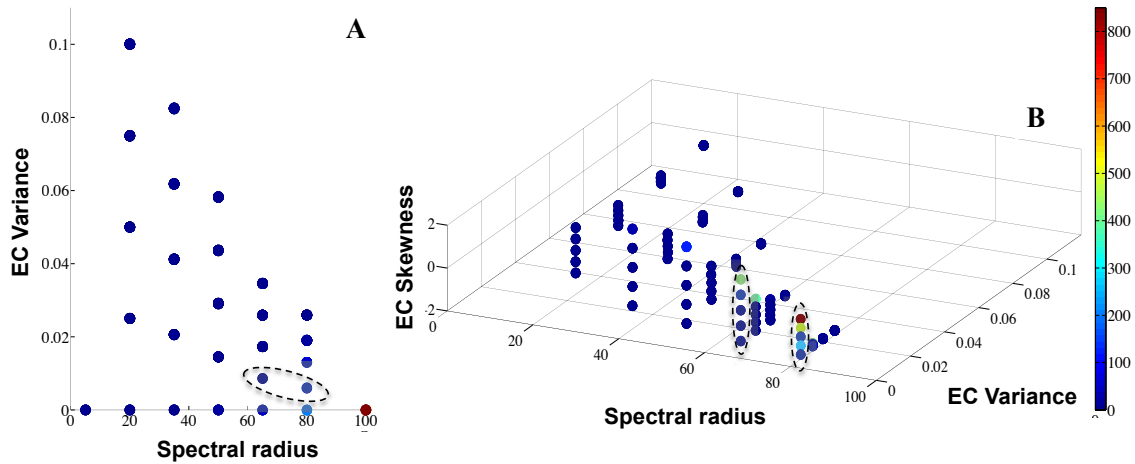

*Figure 6*. Grouping spread outcomes. Color denotes median time to full network occupation. Dark blue coincides with 'fast' spread time and burgundy denotes 'slow' spread. Spread outcomes corresponding to zero EC variance can only be displayed in panel A and not in panel B because skewness is undefined at such points. The area bounded by the dashed oval highlights a region where EC skewness is a strong determinant of spread; this bounded area is titled region I.

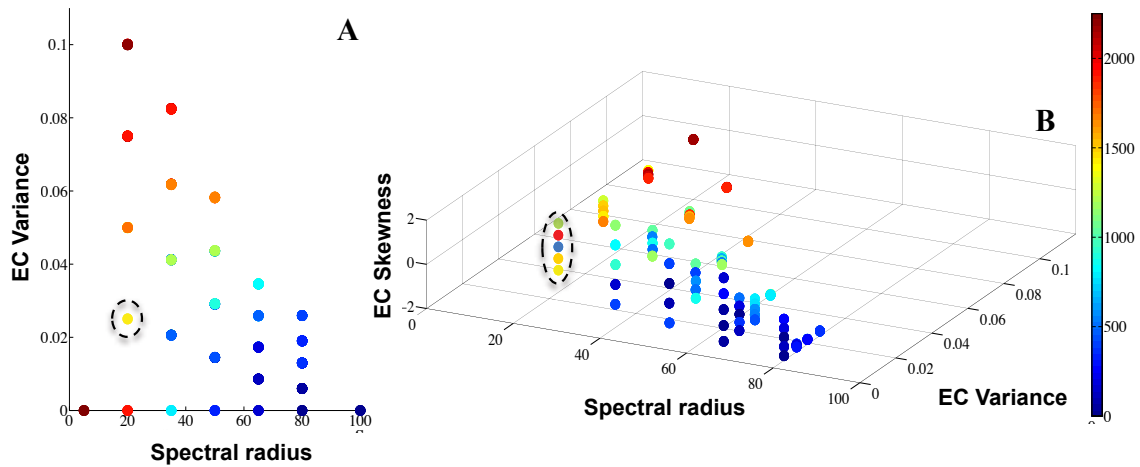

*Figure 7*. Grouping survival outcomes. Color denotes median time to extinction. Dark blue coincides with 'brief' survival and dark red denotes 'long' survival. The area bounded by the dashed oval highlights a region where extinction time is variable and dependent on EC skewness; the area is titled region II.

Persistence outcomes are highly variable when both spectral radius and EC variance are low. We expect high traversability (low spectral radius) to ensure survival, but the relative number of strong contributor nodes, measured by EC skewness, is the deciding factor. On average, networks with a spectral radius of 20km and EC variance of 0.02 induce moderately high persistence times, but if the EC skewness is zero, indicating an equal number of weak and strong contributor nodes, then persistence times plummet (see region *II* in figure 7A-B). Large asymmetry in node contribution strength, which occurs at both ends of the EC skewness spectrum, supports greater persistence times.

Despite the influential role of net contribution strength when determining persistence outcomes for seemingly similar networks, the pattern in region *II* is not uniform across the tri-metric space (figure 7B). Generalizing the effects of EC skewness is difficult, but we find persistence time is longer at one of the extremes. These findings support the notion that for fixed spectral radius and EC variance, one strongly connected node—or the absence of a weakly connected node—may be more important for network cohesion than multiple moderately connected nodes. These results also suggest that there is strong nonlinearity with respect to how skewness interacts with spectral radius and variance; in this case, studying higher moments may play an important role in unpacking and explaining the nonlinearity. Such an extension does not contradict the conclusion that applying the moments concept makes network features straightforward to measure.

**DISCUSSION AND CONCLUSION**

Network metrics and node centralities collapse the high dimensionality of networks into a single dimension. Therefore, no single metric can precisely describe spread or survival on a network. Using multiple metrics in a systematic manner helps retain structural information and describe different network attributes influencing a process occurring on a network. It is possible to negotiate tradeoffs between simple, readily interpretable metrics and the amount information lost through summarization by thinking systematically about how the information from a network is summarized.

      A systematic approach to network measurement begins at the global scale with the most general metric of structure (e.g. a network metric), and then categorizes based on individual-scale heterogeneities (e.g. node centrality scores). We recover the information in node centrality scores with routine formulae for statistical moments. The mathematical dependence between the metrics determines the range of possible network configurations. However, large variation in network configuration does not imply large variation in process outcome. High outcome variability manifests itself in specific regions of the metric space, which vary depending on the process considered. For models of spread, variability in the time until full network occupation occurs only for structures with large spectral radius. In this case, multi-metric analysis shows that even poorly traversable networks can foster quick dispersal depending on the number of strong contributors. For models of survival, high variability in extinction time occurs for structures with low spectral radius and EC variance; highly connected systems do not guarantee long-term persistence.

The interplay between grouped metrics highlights tradeoffs in structural design, which broadens the criteria for network selection. In general, the nature of the process and the layout of the edge weights determine the extent to which structural tradeoffs are feasible. In our ecological example, traversability is not the sole driving force behind long-term persistence and can be substituted by greater node heterogeneity. Structural tradeoffs also extend across multiple processes. We find that most networks promoting persistence also facilitate dispersal, but the converse is not true. Ordered multi-metric analyses do not provide a definitive summary on network dynamics, but help illustrate and understand the complexities in identifying preferred networks (e.g. structures that minimize invasive species spread, maximize survival, or a combination of both). Applying statistical moments does not create new metrics; it brings order to the large set of available networks metrics and facilitates combining them in a logical manner.

**ACKNOWLEDGEMENTS**

D. Shanafelt, B. Morin, and E. Morales contributed helpful discussions about coding structure and statistical analysis. JAB acknowledges support from the Center for Behavior, Institutions and the Environment. KRS acknowledges support from the NSF Alliance for faculty diversity postdoctoral fellowship [NSF Grant DMS-0946431].

# SUPPORTING INFORMATION

## SI Text S1: Proposition 1

*Let $\vec{v}_G$ represent the eigenvector centrality corresponding to a N-patch network ($N \geq 1$) denoted $G$. Let $\bar{v}$ and $var(\vec{v}_G)$ denote the mean and variance of $\vec{v}_G$ respectively. Then there exists a continuous function $g: \bar{v} \to var(\vec{v}_G)$, such that $g$ is bijective.*

*Proof.*

$$\bar{v} = \sum_{i=1}^{N} v_i / N \qquad [S1]$$

$$var(\vec{v}_G) = \frac{1}{N} \sum_{i=1}^{N} (v_i - \bar{v})^2 \qquad [S2]$$

Expanding eqn S2 and using the identity in eqn S1 gives,

$$var(\vec{v}_G) = \frac{1}{N} \sum_{i=1}^{N} (v_i^2 - 2 v_i \bar{v} + \bar{v}^2)$$

$$= \frac{1}{N} \sum_{i=1}^{N} v_i^2 - \frac{2\bar{v}}{N} \sum_{i=1}^{N} v_i + \bar{v}^2$$

$$= \frac{1}{N} \left( \sum_{i=1}^{N} v_i^2 \right) - 2 \bar{v}^2 + \bar{v}^2$$

$$= \frac{1}{N} \left( \sum_{i=1}^{N} v_i^2 \right) - \bar{v}^2 \qquad [S3]$$

Eigenvector centralities are rescaled so the corresponding Euclidean norm equals one, so

$$\sqrt[2]{\sum_{i=1}^{N} v_i^2} = 1 = \sum_{i=1}^{N} v_i^2 \qquad [S4]$$

Substituting eqn S4 into eqn S3 yields,

$$var(\vec{v}_G) = \frac{1}{N} - \bar{v}^2 = g(\bar{v}) \qquad [S5]$$

As $\bar{v}$ can only take on positive values, $var(\vec{v}_G)$ decreases monotonically with greater mean values; thus $g$ is one-to-one and surjective. eqn S5 represents a bijective mapping, $g$, between the mean and variance of the eigenvector centrality.

## SI Text S2: Connecting eigenmetrics and other popular network metrics

*Assumption A1*: We consider a specific class of fully connected networks with corresponding adjacency matrices that are zero-diagonal, nonnegative and symmetric. Without loss of generality, we refer to such a network as network $G$.

*Proposition 2: If (A1) holds, then the spectral radius of network G is a rescaled version of its mean strength that accounts for EC scores.*

*Proof.* The strength of node $i$ in network $G$, indicated as $w_i$, is calculated as the sum of the weights on its connecting edges,

$$w_i = \sum_j w_{i,j} \quad \quad [S6]$$

where $w_{i,j}$ is the weight of the edge connecting nodes $i$ and $j$. The mean strength of network $G$ is the mean of $w_i$ over the $N$ nodes of network $G$,

$$\overline{w} = \frac{\sum_i w_i}{N} = \frac{\sum_i \sum_j w_{i,j}}{N} = \frac{\sum_{i,j} w_{i,j}}{N} \quad \quad [S7]$$

$A_G = \{w_{i,j}\}$ is the adjacency matrix corresponding to network $G$. Let $\vec{v}_G$ represent the eigenvector centrality corresponding to $A_G$. Then one formula for the spectral radius,[1] $\lambda_G$, of $A_G$—provided by Frobenius (Cao 1998)—is,

$$\lambda_G = \frac{\sum_i \sum_j w_{i,j} v_j}{\sum_i v_i} = \frac{\sum_{i,j} w_{i,j} v_j}{\sum_i v_i} \quad \quad [S8]$$

where $v_i \in \vec{v}_G$. Eqn S7 is an arithmetic mean strength that places equal value on each node, eqn S8 is a "rescaled" arithmetic mean strength that takes into account the centrality score of the corresponding node (we use "rescaled" but the more popular term is "weighted", we do not use the latter term in this context because of conflict with its use to describe the connection between nodes). For a rescaling to have properties that the concept of rescaling generally implies, arithmetic means and "rescaled" arithmetic means must be positively correlated.

*Proposition 3: If (A1) holds, then the mean strength and mean clustering coefficient of Network G are related by a constant factor.*

*Proof.* The weighted clustering coefficient (Onnela et al. 2005; MATLAB Brain Connectivity Toolbox] of node $i$ is,

$$c_i = \frac{\sum_{j,k} (\overline{w_{i,j}}\, \overline{w_{j,k}}\, \overline{w_{k,i}})^{1/3}}{\eta_i (\eta_i - 1)}$$

where the weights are rescaled by the largest weight in the entire network, $\overline{w_{i,j}} = w_{ij}/\max(w_{ij})$. $\eta_i$ is the total number of neighbors belonging to node $i$. Since all networks with weighted edges can be considered fully connected, it must be the case that $\eta_i = N - 1$. Thus,

$$c_i = \frac{\sum_{j,k}(w_{i,j} w_{j,k} w_{k,i})^{1/3}}{(N-1)(N-2)\max(w_{i,j})}$$

---

[1] There is more than one formula for calculating the spectral radius of a network. For example, the spectral radius of a star graph can be calculated as $\sqrt{\sum_i w_i^2}$, where each $w_i$ is a distinct edge weight in the network. However, such formulas are specific to networks with certain properties and do not extend to the fully connected graphs we consider. The formula we use fits graphs with more general properties, this is because the Frobenius formula is derived from the basic definition of an eigenvalue.

Using the above equation, the mean clustering coefficient has the form,

$$\bar{c} = \frac{\sum_i c_i}{N} = \frac{\sum_i \frac{\sum_{j,k}(w_{i,j} w_{j,k} w_{k,i})^{1/3}}{(N-1)(N-2)\max(w_{i,j})}}{N}$$

$$= \frac{\sum_i \sum_{j,k}(w_{i,j} w_{j,k} w_{k,i})^{1/3}}{N(N-1)(N-2)\max(w_{i,j})}$$

$$\leq \frac{\sum_i \sum_{j,k} \frac{w_{i,j} + w_{j,k} + w_{k,i}}{3}}{N(N-1)(N-2)\max(w_{i,j})} \quad \text{(by the arithmetic - geometric mean equality)}$$

$$= \frac{\sum_i \sum_{j,k} w_{i,j} + \sum_i \sum_{j,k} w_{j,k} + \sum_i \sum_{j,k} w_{k,i}}{3N(N-1)(N-2)\max(w_{i,j})}$$

$$= \frac{\sum_k \sum_{i,j} w_{i,j} + \sum_i \sum_{j,k} w_{j,k} + \sum_j \sum_{k,i} w_{k,i}}{3N(N-1)(N-2)\max(w_{i,j})}$$

$$= \frac{N \sum_k \bar{w} + N \sum_i \bar{w} + N \sum_j \bar{w}}{3N(N-1)(N-2)\max(w_{i,j})} = \frac{3N^2 \bar{w}}{3N(N-1)(N-2)\max(w_{i,j})} = \frac{N \bar{w}}{(N-1)(N-2)\max(w_{i,j})}$$

Recall $\bar{w}$ is the mean strength. The above derivation shows that a strong upper bound for the mean clustering coefficient exists as a linear function of the mean strength.

*Proposition 4: If (A1) holds, then the mean shortest path length of Network G characterizes a strong lower bound for EC variance.*

*Proof.* The mean shortest path length of network $G$ is written as,

$$l(G) = \frac{1}{N(N-1)} \sum_{i,j \in G} d_{i,j} \quad \text{[S9]}$$

where $d_{ij}$ is the shortest weighted path between nodes $i$ and $j$. Using an alternate formula for EC variance derived in *SI Text S1*, we write,

$$var(\vec{v}_G) = \frac{1}{N} - \bar{v}^2 = \frac{1}{N} - \left(\frac{\sum_{i=1}^N v_i}{N}\right)^2$$

$$= \frac{1}{N} - \left(\frac{\sum_{i,j} w_{i,j} v_j}{N \lambda_G}\right)^2 \quad \text{(Using eqn S8)}$$

$$= \frac{1}{N} - \left(\frac{\sum_j w_j v_j}{N\lambda_G}\right)^2 \quad \text{(Using eqn S6)}$$

$$\geq \frac{1}{N} - \frac{\left(\sum_j w_j^2\right)\left(\sum_j v_j^2\right)}{[N\lambda_G]^2} \quad \text{(Using Cauchy's inequality)}$$

$$= \frac{1}{N} - \frac{\left(\sum_j w_j^2\right)}{[N\lambda_G]^2} \quad \text{(Using eqn S4 in \textit{SI Text S1})}$$

$$\geq \frac{1}{N} - \left(\frac{\sum_j w_j}{N\lambda_G}\right)^2 \quad \text{[S10]}$$

Latora and Marchiori (2001) assert that when nodes $i$ and $j$ are connected, the weight of their adjoining edge is equal to the shortest path between them, i.e. $w_{i,j} = d_{i,j}$. Since we assume each node is connected to every other node in $G$,

$$w_{i,j} = d_{i,j} \text{ for all } i, j \in G,$$

which further implies $\sum_j w_j = \sum_{i,j} w_{i,j} = \sum_{i,j} d_{i,j} = \sum_j d_j$. So inequality S10 becomes,

$$var(\vec{v}_G) \geq \frac{1}{N} - \left(\frac{\sum_j d_j}{N\lambda_G}\right)^2 \quad \text{[S11]}$$

Using the equation for mean shortest path length given eqn S9, rewrite inequality S11 as,

$$var(\vec{v}_G) \geq \frac{1}{N} - \left(\frac{N-1}{\lambda_G} l(G)\right)^2 \quad \text{[S12]}$$

Inequality S12 shows that the mean shortest path length characterizes a strong lower bound for EC variance.

*Proposition 5: If (A1) holds, then the mean shortest path length of Network G is approximately inversely proportional to global efficiency.*

*Proof.* Global efficiency for a network $G$ is written as,

$$E_{glob} = \frac{1}{N(N-1)} \sum_{i \neq j \in G} \frac{1}{d_{i,j}} \quad \text{[S13]}$$

where $d_{ij}$ is the shortest weighted path between nodes $i$ and $j$. Comparing eqns S9 and S13, mean shortest path length is (approximately) inversely proportional to global efficiency (Latora & Marchiori 2001; Fischer et al. 2014; MATLAB Brain Connectivity Toolbox).

*Proposition 6: If (A1) holds, then the global efficiency of Network G characterizes a strong upper bound for local efficiency.*

*Proof.* Latora and Marchiori (2001) write the equation for efficiency of a node $k$ as,

$$E(G_k) = \frac{1}{\eta_k(\eta_k-1)} \sum_{m,n \in G_k} \frac{1}{d_{m,n}} = \frac{1}{(N-1)(N-2)} \sum_{m \neq n \in G_k} \frac{1}{d_{m,n}} \qquad [\text{S14}]$$

where $G_k$ is the subnetwork of neighbors for node $k$ (i.e. node $k$ is omitted from the network, along with its connections). $\eta_k$ is the total number of nodes in subnetwork $G_k$. Since we have a fully connected network, $\eta_k = N - 1$. Local efficiency of network $G$ is the average efficiency of the subnetworks,

$$E_{loc} = \frac{1}{N} \sum_{k \in G} E(G_k) = \frac{1}{N} \sum_{k=1}^{N} E(G_k)$$

$$= \frac{1}{N} \sum_{k=1}^{N} \left[ \frac{1}{(N-1)(N-2)} \sum_{m,n \in G_k} \frac{1}{d_{m,n}} \right] \qquad \text{(Using eqn S14)}$$

$$\leq \frac{1}{N} \sum_{k=1}^{N} \left[ \frac{1}{(N-1)(N-2)} \sum_{m,n \in G} \frac{1}{d_{m,n}} \right] \qquad \text{(Since } G_k \subseteq G\text{)}$$

$$= \frac{1}{N-2} \sum_{k=1}^{N} \left[ \frac{1}{N(N-1)} \sum_{m,n \in G} \frac{1}{d_{m,n}} \right]$$

$$= \frac{1}{N-2} \left( \sum_{k=1}^{N} E_{glob} \right)$$

$$= \frac{N}{N-2} E_{glob}$$

Global efficiency characterizes a weak upper bound for local efficiency.

**SI Text S3: Pseudo code for matrix generating algorithm**
1. Designate the structure of solution matrix, $x^*$.
    a. Assign matrix dimensions (e.g. $N \times N$).
    b. Assign lower and upper bound for matrix elements.
    c. Assign specific spectral radius ($\lambda_{x^*}$), EC variance ($var(\vec{v}_{x^*})$), and EC skewness ($skew(\vec{v}_{x^*})$).

2. Develop constraints for the solution matrix, $x^*$.
    a. Constrain $x^*$ to have zeros in the diagonal entries.
    b. Constrain $x^*$ to be symmetric.
    c. Constrain $x^*$ to be bounded below and above by values designated in (1b).

3. Generate a $N \times N$ initial matrix, $x_0$, with random entries bounded above by maximum weight designated in (1b) and make note of the corresponding spectral radius, EC variance, and EC skewness.

4. Compare each metric designated in (1c) with its counterpart derived from $x_0$. One method of comparison is by taking the absolute value of the differences (e.g. $|\lambda_{x^*} - \lambda_{x_0}|$, $|var(\vec{v}_{x^*}) - var(\vec{v}_{x_0})|$, $|skew(\vec{v}_{x^*}) - skew(\vec{v}_{x_0})|$).

5. While any of the three absolute differences remain greater than some predetermined tolerance level,
    a. Randomize one of the elements of $x_0$ subject to (1b).
    b. Minimize the sum of squared differences between the metrics designated in 1c and their counterparts derived from $x_0$ by perturbing the elements of $x_0$ subject to constraints detailed in (2). Denote this perturbed version of $x_0$, $\widetilde{x_0}$.
    c. Replace the elements of $x_0$ with those of $\widetilde{x_0}$.
    d. Reevaluate absolute differences as per (4).

6. Once all three absolute differences become less than the predetermined tolerance level, the iterative process outlined in (5) stops and the last updated version of the matrix $x_0$ is designated $x^*$.

**SI Text S4: Schedule of Events**
A description of the sequence of agent events for each ecological process is archived online ([www.openabm.org](www.openabm.org)). The ABM birth, death, and dispersal events are stochastic and occur if a randomly chosen value from the random uniform distribution on the unit interval ([0,1]) is less than the corresponding rate of event occurrence. We simulate 400 realizations per network configuration for each ecological process. Each ecological scenario follows a similar sequence of actions briefly described below and illustrated in Fig. S1:

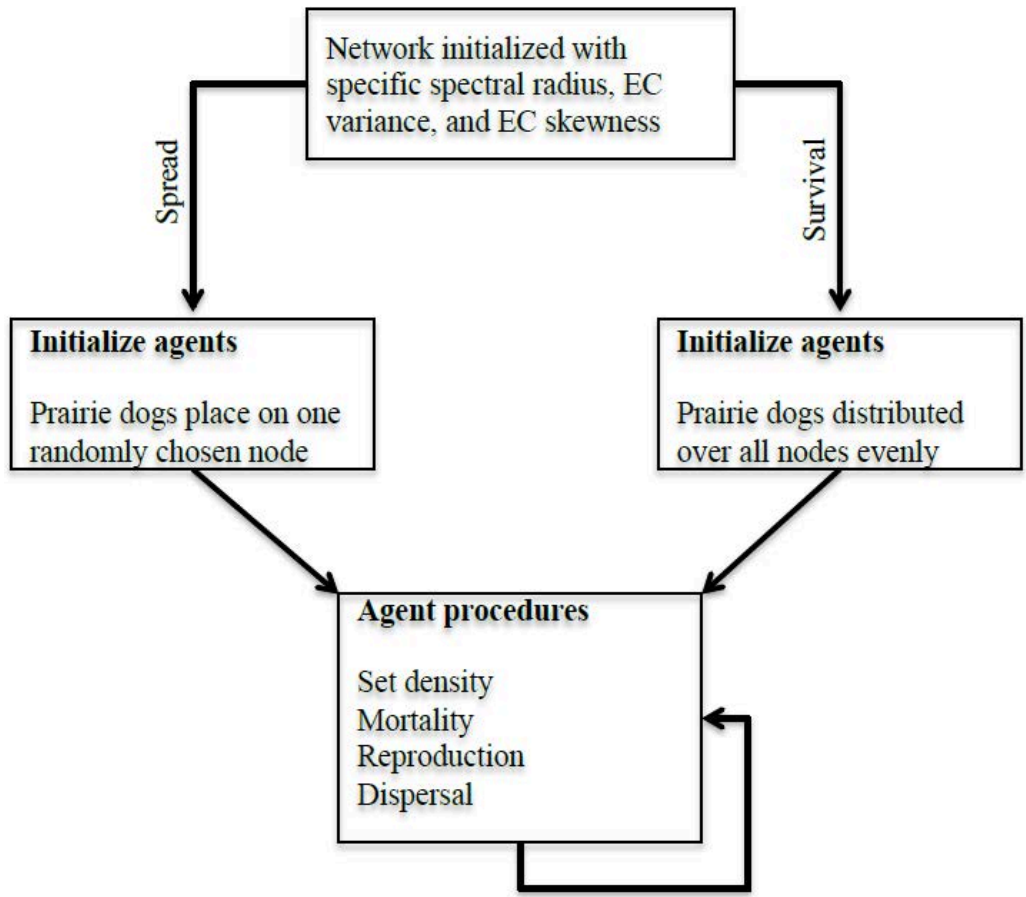

**Figure S1.** *Schedule of events as coded in the agent based model*

1) Landscape initialized with a habitat network ($G$) possessing a specified spectral radius ($\lambda_G$), EC variance ($var(\vec{v}_G)$), and EC skewness ($skew(\vec{v}_G)$).

2) Initialization of agents
    a. *Spread* – A population of $N_x$ Prairie dogs initially placed on one randomly chosen patch.
    b. *Survival* – $N_x$ prairie dogs initially placed on each patch.

3) Internalization of local information
    a. *Spread* - Each prairie dog on patch *i* internalizes local information by counting the number of agents on its patch (including itself) and determining the population density (i.e. $D_{x,i}$). We assume agents do not update these values until the beginning of the next time-step so as to create the idea that agents actions, especially density-dependent events, occur in some simultaneous fashion and no single agent receives the most current information.
    b. *Survival* – Same as above.

4) Individual stochastic events

a. *Spread* - During a time-step, prairie dog natural mortality events occur first. Then surviving prairie dogs may reproduce. Lastly, dispersal events are calculated. Dispersal mortality occurs if prey dispersal is unsuccessful or if intra-species competition is too great on target patch.
   b. *Survival* – Same as above.

5) Stop conditions
   a. *Spread* - If prairie dogs have not reached the last uninhabited patch, process is repeated from step 3.
   b. *Survival* - If prairie dogs still exist on the landscape, process is repeated from step 3.

**SI Text S5: LOWESS Smoothing**
Locally weighted scatterplot smoothing (i.e. LOWESS smoothing) methods fit a low-degree polynomial regression to a subset of the data derived from simulation. The LOWESS method gives higher weights to points nearby and lower weights to points further away from the point where the dependent variable is estimated given the independent variable (Cleveland 1979; Cleveland & Devlin 1988). The weights given to distance between points of the independent variable are assigned according to the following function:

$$w(x) = \begin{cases} (1-|x|^3)^3 & \text{for } |x| < 1 \\ 0 & \text{for } |x| \geq 1 \end{cases}$$

[S6]

**SI Text S6: Describing Supplementary on Sensitivity Analysis**
We use this section to help readers understand the information in the supplementary files titled 'spread_sensitivity_tables.txt' and 'survival_sensitivity_tables.txt'. The '.txt' files provide results on how a ten percent increase/decrease in each default population parameter (see Table 1) changes the statistical metrics (e.g. $R^2$) of a given regression model in the spread and survival scenario. Each '.txt' file contains 14 tables, each table has a label above it with the name of the perturbed parameter and the corresponding perturbed value. The contents of each table include a list of regression models and their corresponding statistics. The names given to the network metrics (i.e. the independent variables) differ from those in the main manuscript, we translate them here; eigen – spectral radius, vareig – EC Variance, skeig – EC Skewness. We use log-likelihood values in the '.txt' files and ∆AIC values in the main manuscript; lower ∆AIC values coincide with higher log-likelihood values. The main result is not altered by the change in statistical measures; the best regression models (with relatively high $R^2$, low ∆AIC values, and high log-likelihood values) still use all three network metrics.